\input amstex
\documentstyle {amsppt}

\topmatter
\title
   Internal Turing Machines
\endtitle
\author Ken Loo \endauthor
\abstract
  Using nonstandard analysis, we will
  extend the classical Turing machines
  into the internal Turing machines. 
  The internal Turing machines have
  the capability to work with infinite 
  ($*$-finite) number of bits while
  keeping the finite combinatoric structures
  of the classical Turing machines.  
  We will show the following.
  The internal deterministic Turing machines can do in
  $*$-polynomial time what
  a classical deterministic
  Turing machine can do in an arbitrary finite amount of
  time. 
  Given an element of $<{}^*M;{}^*x>\in Halt$ (more precisely,
  the $*$-embedding of $Halt$),
  there is an internal deterministic Turing machine
  which will take $<{}^*M;{}^*x>$ as input and  halt in the $"yes"$ state,
  and for $<{}^*M;{}^*x> \notin Halt$, the internal deterministic
  Turing machine will halt in the $"no"$ state.  
  The language ${}^*HALT$ can not be decided by the internal 
  deterministic Turing machines.  The internal deterministic 
  Turing machines
  can be viewed as the asymptotic behavior of 
  finite precision approximation to real number computations.
  It is possible to use the internal probabilistic 
  Turing machines to 
  simulate finite state quantum mechanics with infinite
  precision.  This simulation suggests that no information
  can be transmitted instantaneously and at the same time,
  the Turing machine model can simulate instantaneous
  collapse of the wave function. 
  The internal deterministic Turing machines are powerful,
  but if $P \neq NP$,
  then there are internal problems which the internal 
  deterministic Turing machines
  can solve but not in $*$-polynomial
  time. 
\endabstract

\address P.O. Box 9160, Portland, OR. 97207
\endaddress

\email look\@sdf.lonestar.org \endemail
\leftheadtext{Internal Turing Machines}
\rightheadtext{Internal Turing Machines}

\endtopmatter

\document
\subhead{\bf 1. Introduction}
\endsubhead
   Nonstandard analysis is well known for its ability
   to rigorously work with infinites and infinitesimals 
   in finite discrete combinatoric manner.  We will use
   this ability to extend the classical Turing machines
   to the internal Turing machines.  The outcome of this
   extension can simply be summarized as follows.  Allow the
   classical Turing machines to work with an infinite
   number of bits in infinite amount of time steps in such
   a way that it keeps the finite combinatoric structures of
   the classical machines.  
   
   The internal Turing machine model is a computational model
   that is much more powerful than the classical Turing
   machine model.  As of today, we are not aware
   of any evidence that the internal Turing machine
   model is a reasonable model of computation for the
   universe that we live in, but for the following reasons,
   we believe that this model is worth studying.  The
   classical Turing machines are embedded in the
   internal ones.  Hence, from a purely mathematical point of 
   view, the classical Turing machines live in a much
   bigger universe.  In mathematics, results are often obtained
   by embedding objects into a bigger universe.
   The internal Turing machines are
   capable of performing computations that are infinitesimally
   close to real variable computations.  This could be
   useful as a bridge between finite discrete computations,
   experimental science, and continuous variable physical
   modeling.  The internal probabilistic Turing machine model 
   can simulate finite state quantum mechanics in such
   a way that it can model instantaneous collapse
   of the wave function and suggests the non-allowance of 
   instantaneous transmission of information.  
   This suggests that the classical 
   definition of deterministic Turing machine might be
   more fundamental than previously thought (more on this
   later).  The internal Turing machines are much more
   powerful than the classical ones, but at the same time,
   they have the same types of limitations as the classical
   ones.  For the philosophers and theologians, 
   this suggests that 
   our computational power was created in the image of
   more godly computers which are infinitely more powerful
   than what we can do and at the same time, have the same
   types of limitations as we do.  
    
   The structure of this paper is as follows.  We will
   assume that the reader is familiar with the basic
   notions of nonstandard analysis, complexity theory,
   and quantum physics but we will try to make the
   paper self contained.  In section 2, we will give
   a overview of nonstandard analysis.  The main
   goal is familiarize the reader with the $*$-transform. 
   For more details on nonstandard analysis, we refer
   the reader to [3], [5], [7], and [9].  
   In section 3, we will give 
   a rigorous definition of the classical 
   deterministic Turing machines
   in terms of sets and functions.  This is needed
   to apply the superstructure methods of nonstandard
   analysis.  After this rigorous definition, we will
   extend the classical definition to the internal
   Turing machines.  For the rest of section 3,
   we will take a look at what the internal deterministic
   Turing machines can and can not do.  Namely, we
   will show that the internal Turing machines can do
   in $*$-polynomial time what the classical Turing
   machines can do in finite amount of time.  There
   exists an internal deterministic Turing machines 
   that will answer $"yes"$ on
   input from ${}^\sigma HALT$, the $*$-embedding of 
   $HALT$. On the other hand, the internal
   deterministic Turing machines can not decide ${}^*HALT$.   
   In section 4, we will show that the internal
   deterministic Turing machines can be viewed as the asymptotic
   behavior of the real computation model proposed
   in [6].  In section 5, we will assume that nature
   uses internal Turing machines to compute physical evolutions,
   and we will simulate finite state quantum mechanics with
   the internal probabilistic Turing machines.   
   We will see that the simulation suggests that
   the classical Turing machine model might be
   more fundamental than previously thought. 
   In section 6, we extend the classical nondeterministic
   Turing machines to internal nondeterministic 
   Turing machines.  We will show that even though
   the internal Turing machines are much more powerful
   than the classical ones, they have the same types
   of limitations that the classical Turing machines
   have.

\subhead{\bf 2. Nonstandard Analysis}
\endsubhead  We begin by giving a quick overview of 
nonnstandard analysis.  For more details on
nonstandard analysis, see [3], [5], [7], and [9].

\proclaim{\bf Theorem 1} There exists a finite additive
measure $m$ over $\Bbb N$, the positive integers, such that
the following three properties hold.
For all $A\subset \Bbb N$, $m(A)$ is defined and it is
either $0$ or $1$, $m(\Bbb N) = 1$, and    
$m(A) = 0$ for all finite $A$.   
\endproclaim
\demo{Proof} See [3]. \qed
\enddemo

We now use the measure to construct the nonstandard reals. 
Let  
$$
  S = \left\{\left\{a_n\right\}| a_n\in\Bbb R, n\in\Bbb N\right\} ,\tag2.1
$$
be the collection of all infinite sequences of
real numbers.
Define an equivalent relations on $S$ as follows.
For any two sequences 
$\left\{a_n\right\}$ and $\left\{b_n\right\}$,
$\left\{a_n\right\}\sim \left\{b_n\right\}$ iff
$m\left\{n| a_n = b_n\right\} = 1$.
The nonstandard reals is defined as ${}^*\Bbb R = S/\sim$.
For any two equivalence classes 
$<a_n>$ and $<b_n>$ with
representative sequences  
$\left\{a_n\right\}$ and $\left\{b_n\right\}$,
numerical operations like addition and multiplication are defined by
$<a_n> + <b_n> = 
<a_n + b_n>$, and    
$<a_n> * <b_n> = 
  <a_n * b_n>$.  Orderings like $\leq$ is defined
by 
$<a_n> \leq <b_n>$ iff 
$m\left\{n| a_n \leq b_n\right\} = 1$.  It can be 
shown that these operations are independent of
the representatives.  It can also 
be shown that ${}^*\Bbb R$ is an ordered
field with zero element $0 = <0,0,\dots , 0,\dots>$
and unit element $1 = <1,1,\dots , 1,\dots>$.
Finally, for $a\in\Bbb R$, the mapping 
$a \to <a, a, \dots, a, \dots>$ is an order preserving
homomorphism embedding $\Bbb R$ into ${}^*\Bbb R$.

The nonstandard reals contain a copy of $\Bbb R$
and elements that are not in $\Bbb R$.  Namely,
it contains the infinites, the infinitesimals,
and the reals plus the infinitesimals.
Consider the equivalence class 
$\omega = <1, 2, 3, \dots>$.  For $a\in\Bbb R$,
its embedding in ${}^*\Bbb R$ is given by
$<a, a, \dots>$.  Further, 
$\left\{\lceil a\rceil, \lceil a\rceil + 1, \dots \right\} = 
\left\{n| a \leq n, n\in\Bbb N\right\}$
and $m\left\{n| a \leq  n, n\in\Bbb N\right\} = 1$.  
The last result comes from
$$\align
   {}&1 = m(\Bbb N) = 
      m\left(\left\{\lceil a\rceil, \lceil a\rceil + 1 , \dots \right\}\cup
       \left\{1, \dots \lceil a\rceil - 1 \right\}\right) =  \tag2.2 \\ 
    {}&m\left(\left\{\lceil a\rceil , \lceil a\rceil + 1 , 
        \dots \right\}\right) +
      m\left(\left\{1, \dots \lceil a\rceil - 1\right\}\right) =
     m\left(\left\{\lceil a\rceil , \lceil a\rceil + 1, \dots 
     \right\}\right) + 0 . 
\endalign
$$
This shows that for any $a\in\Bbb R$,
$<a,a,\dots >$ is less than $<1,2,3,\dots>$.  The
nonstandard real numbers of this type are the infinite
nonstandards.  For our purpose, we will mainly be
interested in the nonstandard infinite integers.    
These are nonstandard integers of the form
$<b_n>, b_n\in\Bbb Z$ for all $n\in\Bbb N$, and
for all $a\in\Bbb Z$, $|<a,a,\dots>| \leq
|<b_n>|$, where 
$|<a_n>| \leq |<b_n>|$ is defined by
$$
   |<a_n>| \leq |<b_n>| \quad\text{iff}\quad
   m(\left\{n||a_n| \leq |b_n|\right\}) = 1. \tag2.3
$$
At the other end of the spectrum are the infinitesimals.
Consider the equivalence class $\delta = <1, 1/2, 1/3, \dots ,1/n, \dots>$.
For any $a\in\Bbb R - 0$, 
$$
  <0,0,\dots> \text{ }<  \delta < \text{ }|<a, a, \dots>| . \tag2.4
$$  
Hence, $\delta$ is smaller than $|a|$ for all $a\in\Bbb R - 0$.
The nonstandard number $\delta$ is an infinitesimal.  Both
the infinites and the infinitesimals can be positive or negative,
and the nonstandard number 0 is the only infinitesimal in $\Bbb R$.
An element $x\in {}^*\Bbb R$ is called finite if there exists
a positive real number $a\in\Bbb R$ such that
$-a <  x < a$.  It can be shown that for any finite
$x\in {}^*\Bbb R$, there is a real $a\in\Bbb R$ 
and an infinitesimal $\delta\in {}^*\Bbb R$ such that
$x = a + \delta$.  For any $x,y\in {}^*\Bbb R$, we
write $x\approx y$ to denote that $x = y + \delta$,
where $\delta$ is an infinitesimal.  In other-words,
we say that $x$ and $y$ are infinitesimally close 
to each other.  For any finite nonstandard $x = a + \delta$,
we denote the standard part of $x$ by $st(x) = a$.  
It can be shown that if the sequence $\left\{a_n\right\}$
converges to $a$, then $a \approx <a_n>$ or 
$a = st(<a_n>)$.
Finally, the set of all $x$ such that $st(x) = a$ is called
the monad of $a$, i.e., the monad of $a\in\Bbb R$ is the
set $a + \delta$ where $\delta$ is an infinitesimal.

We now look at the sets of ${}^*\Bbb R$.
A sequence of subsets $\left\{A_n\right\}$ of $\Bbb R$
defines a subset $<A_n>$ of ${}^*\Bbb R$ by
$<x_n> \in <A_n>$ iff 
$m(\left\{n|x_n\in A_n\right\}) = 1$.  Any subset of 
${}^*\Bbb R$ that can be obtained this way is called
internal.  For any subset $A$ of $\Bbb R$, the internal
subset ${}^*A = <A>$ is called the nonstandard version of $A$.
An internal set is called standard if it is of the form
${}^*A$.  For any subset $A$ of 
$\Bbb R$, $A\subseteq {}^*A$ with equality hold if
and only if $A$ is finite.  The ${}^*$ operation
is called the $*-$transform.  Hence, the $*-$transform
of an infinite set is a proper extension of the set.
An internal set $A = <A_n>$ is called hyperfinite
if almost all of the $A_n$'s are finite; the set $A$
has internal cardinality $|A| = <|A_n|>$ where
$|A_n|$ is the number of elements in $A_n$.
There are sets of ${}^*\Bbb R$ that
are not internal.  
The set of infinitesimals of ${}^*\Bbb R$ is not an
internal set.  A set that is not internal is called
external.  

\proclaim{\bf Example 1}
Let $N = <N_n> \in {}^*\Bbb N$ be an infinite element
of nonstandard natural numbers.  The set
$T = \left\{0, 1/N, 2/N, \dots , (N - 1)/N, 1\right\}$
should have internal cardinality $N + 1$ (notice that
$N$ is an infinite integer).  Since $N = <N_n>$,
$T = <T_n>$, where 
$T_n = \left\{0, 1/N_n, 2/N_n, \dots ,1\right\}$.  
The cardinality of $T$ is then $|T| = <|T_n|>
= <N_n + 1> = N + 1$.  The point here is that
the set $T$ has infinite cardinality, but
we can treat the combinatorics as if it were finite. 
Properties like this is better illustrated with the 
*-transform theorem below.
\endproclaim  

After having defined sets, we now define functions.
Let $f_n: A_n\to B_n$ be a sequence of functions.
The internal function $<f_n>: <A_n>\to <B_n>$
is defined by $<f_n>{<x_n>} = <f_n(x_n)>$.  
Any function that can be obtained this way is
called an internal function.
If $f: A \to B$, then the function 
${}^*f = <f>: <A> \to <B>$
is called the nonstandard version of $f$.
An internal function which can be obtained
this way is called standard.  
A characterization of the internal cardinality
of an internal set $A$ is that there exist an
internal bijection $f:\left\{1,2,\dots ,N\right\}
\to A$.
   
To make the process of working with nonstandard analysis
less cumbersome, we introduce superstructures and the 
$*$-transform on superstructures.  This allows us to
work with nonstandard analysis described above in one big
scoop.  For our purpose,
we will mainly be interested in functions and relations.
Functions and relations can be described in terms of set theory.
Given a set $S$, an $n$-ary relation $P$ on $S$ is a subset of 
$S^n = S\times S \dots \times S$.  For 
$<a_1, a_2, \dots , a_n> \in P \subseteq S^n$, 
The domain of P consists of those elements 
$<a_1, a_2, \dots , a_{n-1}>$ 
of $S^{n-1}$ such that there exists an $a$ in $S$ with
$<a_1, a_2, \dots , a_{n-1}, a> \in P \subseteq S^n$. 
The set of all such $a$'s is the range of $P$.
A relation $P$ is a function if whenever
$$
  <a_1, a_2, \dots , a_{n-1}, a> =  
 <a_1, a_2, \dots , a_{n-1}, b>,  \tag2.5
$$ then $a = b$.
We now define superstructures.  For a set 
$X$, $\Cal P\left(X\right)$ is the power set
or the set of all subsets (including $\emptyset$) of $X$.     
The $n$-th cumulative power set of $X$ is defined 
recursively by
$$
  V_0(X) = X, V_{n+1} = V_n(X)\cup \Cal P(V_n(X)) . \tag2.6
$$  
The superstructure over $X$ is the set
$$
  V(X) = \bigcup_{n=0}^{\infty} V_n(X) . \tag2.7
$$

The symbols of the language for 
$V(X)$ consist of the following.
The connective symbols $\neg$, $\vee$,
$\wedge$, $\rightarrow$, and $\leftrightarrow$ 
will be interpreted as "not", "or", 
"and", "implies", and "if and only if" 
(we will occasionally overload the symbol
$\rightarrow$ as function domain to range notation).
The quantifier symbols $\forall$, and $\exists$
will be interpreted as "for all" and "there exists".
The symbols $[,],(,),$ and $<,>$ will be used
for bracketing.  At least one symbol $a$ for each
element $a\in V(X)$ (for notation convenience,
we have used the same notation for both the element
and its symbol).  A countable collection of
symbols like $x,y,z, \dots$ to  be used
as variables.  The symbol $=$ will be interpreted
as "equal", and the symbol $\in$ will be interpreted
as "an element of".  

A formula of the language is built up inductively as follows.
If $x_1, x_2, \dots , x_n, x$, and $y$ are either constants
or variables, the expressions $x\in y$, 
$<x_1, \dots ,x_n>\in y$,
$<x_1, \dots ,x_n> = y$, 
$<<x_1, \dots ,x_n>,x> \in y$, and 
$<<x_1, \dots ,x_n>,x> = y$ are formulas
called atomic formulas.  If $\Phi$ and
$\Psi$ are formulas, then so are
$\neg\Phi$, $\Phi \vee \Psi$, $\Phi \wedge \Psi$,
$\Phi\rightarrow\Psi$, and $\Phi\leftrightarrow\Psi$.
If $x$ is a variable symbol, and $y$ is either a
variable or a constant symbol,
and $\Phi$ is a formula which does not already contain an
expression of the form 
$(\forall x\in z)$ or 
$(\exists x\in z)$ (with the same variable $x$),
then $(\forall x\in y)\Phi$ and 
$(\exists x\in y)\Phi$ are formulas.
A variable occurs in the scope of a quantifier
if whenever a variable $x$ occurs in $\Phi$, then
$x$ is contained in a formula $\Psi$ which
occurs in $\Phi$ in the form
$(\forall x\in z)\Psi$ or
$(\exists x\in z)\Psi$; it is then said to
be bounded, and otherwise it is called free.  A
sentence is a formula in which all
variables are bounded.

The decision of true or false of a sentence
$\Phi$ in the language of $V(X)$ is as follows.
The atomic sentences $a\in b$,
$<a_1, \dots , a_n>\in b$, 
$<<a_1, \dots , a_n>,c>\in b$, 
$a = b$,
$<a_1, \dots , a_n> =  b$, and 
$<<a_1, \dots , a_n>,c> =  b$ are
true in $V(X)$ if the entity corresponding
to its name is an element of or identical to $b$.
If $\Phi$ and $\Psi$ are sentences, then
$\neg \Phi$ is true if $\Phi$ is not true;
$\Phi \wedge \Psi$ is true if both $\Phi$ and
$\Psi$ are true;
$\Phi \vee \Psi$ is true if at least one of $\Phi$ or
$\Psi$ is true; 
$\Psi \rightarrow \Phi$ is true
if either $\Phi$ is true or $\Psi$ is not true;
$\Psi \leftrightarrow \Phi$ is true
if $\Psi$ and $\Phi$ are either both true
or both not true.
The expression $(\forall x \in b)\Phi$ is true
if for all entities $a\in b$, when the symbol
corresponding to the entity $a$ is substituted for $x$
in $\Phi$, the resulting formula $\Phi(a)$ is true.
The expression $(\exists x \in b)\Phi$ is true
if there is an entity $a\in b$ such that 
$\Phi(a)$ is true.

\proclaim{\bf Example 2} Let us denote the
set of polynomials with coefficients in 
$\Bbb N\cup\left\{0\right\} \equiv \bar\Bbb N$
and domain 
$\bar\Bbb N$
by $\bar\Bbb NPOLY$, 
then the following statements
are true
$$\align
  {}&(\forall p\in\bar\Bbb NPOLY)
  (\exists n\in \bar\Bbb N)\big[
  (\exists a_0\in \bar\Bbb N)
  (\exists a_1\in \bar\Bbb N)
   \dots
  (\exists a_n\in \bar\Bbb N) \tag2.8\\
  {}&[(<x,y>\in p \rightarrow x\in\bar\Bbb N) \wedge
      (<x,y>\in p \rightarrow y = a_0x^0 + a_1x + \dots + a_nx^n>) 
        ] \big], \\
  {}&(\forall n\in\bar\Bbb N)\bigg[
     (\forall a_0\in \bar\Bbb N)
      (\forall a_1\in \bar\Bbb N)
         \dots
       (\forall a_n\in \bar\Bbb N)(\exists p\in\bar\Bbb NPOLY)\\
  {}&\big[(z\in p \rightarrow z = <x,y>) \wedge 
      (<x,y>\in p \rightarrow x\in\bar\Bbb N) \wedge \\
  {}& (<x,y>\in p \rightarrow y = a_0x^0 + a_1x + \dots + a_nx^n>) 
         \big]\bigg],
\endalign
$$
where we have used the set theoretic notation for functions.
This is just a fancy way of saying that
$p$ is a polynomial with coefficient in $\bar\Bbb N$
if and only if 
$$
   p(x) = \sum_{i=0}^n a_ix^i : \bar\Bbb N \rightarrow
    range(p), a_i\in\bar\Bbb N . \tag2.9 
$$
\endproclaim

The $*$-transform 
of superstructures together with the superstructures'
language and
true or false assignment allow us to efficiently 
apply the methods of nonstandard analysis. 
The injection 
$*:V(\Bbb R) \to V({}^*\Bbb R)$ has the following
properties.  It preserves basic set operations,
${}^*\emptyset = \emptyset$, 
${}^*(A\cup B) = {}^*A\cup {}^*B$, 
${}^*(A - B) = {}^*A - {}^*B$, 
${}^*(A\cap B) = {}^*A\cap {}^*B$, 
${}^*(A\times B) = {}^*A\times {}^*B$, 
${}^*<a_1, a_2, \dots ,a_n> = 
 <{}^*a_1, {}^*a_2, \dots ,{}^*a_n>$. 
It preserves domain and range of relations,
$dom({}^*\Phi) = {}^*dom(\Phi)$,
$rng({}^*\Phi) = {}^*rng(\Phi)$.
It preserves standard definition of sets,
${}^*\left\{(x,y):x\in y\in A\right\} =
\left\{(z,w):z\in w\in {}^*A\right\}$.
It produces a proper extension: 
${}^{\sigma}A \subseteq {}^*A$, 
where 
${}^{\sigma}A = \left\{{}^*x : x\in A\right\}$ is
the $*$-embedding of $A$, and
equality
holds if and only
if $A$ is a finite set.
For all $a\in\Bbb R$ implies ${}^*a\in {}^*\Bbb R$,
and $a\in\Bbb R$ implies ${}^*a = a$.
If $a\in V_{n+1}(\Bbb R) - V_{n}(\Bbb R)$,
${}^*a\in V_{n+1}({}^*\Bbb R) - V_{n}({}^*\Bbb R)$.
If $a\in {}^*V_n(\Bbb R), n\geq 1$
and $b\in a$, then
$b\in {}^*V_{n-1}(\Bbb R)$.
For any sentence $\Phi$ in the language of
$V(\Bbb R)$, ${}^*\Phi$ is the sentence
in $V({}^*\Bbb R)$ obtained by replacing all
"constants" in $\Phi$ by ${}^*$"constants".
For any sentence $\Phi$ in the language of
$V(\Bbb R)$, $\Phi$ is true in $V(\Bbb R)$
if and only if ${}^*\Phi$ is true in
the language of $V({}^*\Bbb R)$.  
Any entity $A$ of the form $A = {}^*B$ is called
standard.  Any element $a\in {}^*B$ is called internal.
An entity that is not internal (standard entities are
internal) is called external.  A sentence or formula
$\Phi$ in the language of $V\left({}^*\Bbb R\right)$
is called either internal or standard if the constants
in $\Phi$ are names of internal or standard entities.
A sentence which is not internal is called external.
 
\proclaim{\bf Example 3} The $*$-transform of the
sentences in example 2 are
$$\align
  {}&(\forall p\in {}^*\bar\Bbb NPOLY)
  (\exists n\in {}^*\bar\Bbb N)\big[
  (\exists a_0\in {}^*\bar\Bbb N)
  (\exists a_1\in {}^*\bar\Bbb N)
   \dots
  (\exists a_n\in {}^*\bar\Bbb N) \tag2.10\\
  {}&[(<x,y>\in p \rightarrow x\in {}^*\bar\Bbb N) \wedge
      (<x,y>\in p \rightarrow y = a_0x^0 + a_1x + \dots + a_nx^n>)
        ] \big], \\
  {}&(\forall n\in {}^*\bar\Bbb N)\bigg[
      (\forall a_0\in {}^*\bar\Bbb N)
      (\forall a_1\in {}^*\bar\Bbb N)
         \dots
       (\forall a_n\in {}^*\bar\Bbb N)
        (\exists p\in{}^*\bar\Bbb NPOLY)\\
  {}&\big[(z\in p \rightarrow z = <x,y>) \wedge
      (<x,y>\in p \rightarrow x \in {}^*\bar\Bbb N) \wedge \\
  {}& (<x,y>\in p \rightarrow y = a_0x^0 + a_1x + \dots + a_nx^n>)
         \big]\bigg].
\endalign
$$
Since the original sentences are true, their $*$-transforms
are also true.  This says that ${}^*\bar\Bbb NPOLY$ 
consists of elements of the form
$a_0 + a_1x + \dots + a_nx^n$ where
the coefficients and $n$ are in $0\cup {}^*\Bbb N$.
Elements $p \in {}^*\bar\Bbb NPOLY$
are internal polynomials.  If the coefficients and
$n$ are in ${}^{\sigma}\bar \Bbb N = \bar\Bbb N$,
the $*$-embedding of $\bar\Bbb N$, then the polynomial
is the $*$-embedding of a "classical" polynomial and
it is a standard entity.  The extension consists
of those internal polynomials which some or all of
the coefficients and $n$ are infinite number(s) in 
${}^*\bar\Bbb N = 0\cup {}^*\Bbb N$.   
\endproclaim

For our purposes, the nonstandard extension allows us
to work with infinite integers and to keep the
combinatoric structures of the standard finite integers.
This is what gives life to the internal Turing machines.

\subhead{\bf 3. Deterministic Turing Machines}
\endsubhead
We start this section with a 
semi-formally description of a k-tape deterministic 
Turing machine,
this definition is taken from [8].
A $k$ tape Turing machine, 
where $k\in\Bbb N$, is a quadruple
$$
   M = <K, \Sigma , \delta , s>, \tag3.1
$$
where $K$ is a countable
set of states; $s\in K$ is the initial state; 
$\Sigma$ is a countable set of symbols; $\Sigma$
contains the blank and the first symbol 
$\sqcup$ and $\vartriangleright$; 
$\delta$ is a transition function which
maps $K\times\Sigma^k$ to 
$(K\cup\left\{h, "yes", "no"\right\})
 \times\left(\Sigma\times\left\{\leftarrow, 
  \rightarrow, -\right\}\right)^k$, where 
$h$ (halting state), $"yes"$ (accepting state),
$"no"$ (rejecting state), $\leftarrow$ (left),
$\rightarrow$ (right), and $-$ (stay) are not 
in $K\cup\Sigma$.  Further, the transition
function $\delta$ has the following property.
If $\delta\left(q, \sigma_1, \dots ,\sigma_k\right) =
 \left(p, \rho_1, D_1, \dots , \rho_k, D_k\right)$,
and $\sigma_i = \vartriangleright$, then
$\rho_i = \vartriangleright$ and $D_i = \rightarrow$.
Intuitively, if the machine $M$ is in state $q$ and
the cursor of the first tape reads $\sigma_1$,
the second tape reads $\sigma_2$, and so on, 
then the next state will be $p$, the cursor of 
the first tape will write $\rho_1$ over $\sigma_1$ and
then move in the direction $D_1$,
of the second tape will write $\rho_2$ over $\sigma_2$,
then move in the direction $D_2$, and so on.  Initially,
all tapes start with a $\vartriangleright$ and this
symbol can not be overwritten.  Further, the  
cursors can not move to the left of the starting point 
$\vartriangleright$ of the corresponding tape.
The $k$-string Turing machine starts its computation
in the configuration
$(s; \vartriangleright, x; 
\vartriangleright, \sqcup;
\dots ;
\vartriangleright, \sqcup)$,
where $\sqcup$ is the blank symbol,
and $x\in (\Sigma - \sqcup)^*$, where
$(\Sigma - \sqcup)^*$ denotes the set
of all finite strings from 
$\Sigma - \sqcup$.
If the machine $M$ reaches state $h$, $"yes"$,
or $"no"$, then the
machine is said to have halted,
accepted the input, or rejected the input.
In the case that the Turing machine computes
functions, the output $y$ is read from the last tape.
This can be denoted by 
$M\left(x\right) = <"yes", y>, 
<"no", y>,$ or $<"h", y>$.  If the
machine computes forever, then it is denoted by
$M(x) = \nearrow$.

Let us divert for a moment to discuss the issue
of time in the above definition.  
The definition of the $k$ tape Turing machine
allows the simultaneous reading or writing
of the tapes at each time step.  What is of 
interest is that the definition does not speak
of the time required for the tape heads
to communicate with the part of the machine
that computes the transition function.   
The definition in fact allows instantaneous
transmission of information at this 
communication junction, which violates
Einstein's special theory of relativity.
In section 5, we will use internal probabilistic
Turing machines to simulate
finite state quantum mechanics.
The time issue just described
allows the internal probabilistic Turing machines 
to model collapse of the wave function.  
On the other hand,
section 5 also suggests that 
information can not be
instantaneously transmitted because given an 
input, it
takes time for the internal Turing machines
to compute the output.   This suggests that
the $k$ tape Turing machines could be
more fundamental than previously thought.

Now back to internal Turing machines.
In order to apply nonstandard analysis to the above, we must
write down the definition of the Turing machine in terms
of sets, functions and relations. 
Let $x$ and
$\delta$ be
the input to the Turing machine and 
the transition function as previously defined.  
We recursively define a sequence of tape states as follows.
For $n\in\bar\Bbb N$, where $\bar\Bbb N$ was defined in section 2,
and for $1\leq i \leq k$,
$$\align
   {}&Tape_i^n: \bar\Bbb N \to \Sigma,  \tag3.2 \\
   {}&Cursor_i: \bar\Bbb N \to \bar\Bbb N, \\ 
   {}&State: \bar\Bbb N \to K \cup \left\{"yes", "no", h\right\}. \\ 
\endalign
$$  
Here,
$Tape_i^n(m)$ is the symbol of the $m^{th}$ cell
of the $i^{th}$ tape at the $n^{th}$ time step,  
$Cursor_i(n)$ is the position of the $i^{th}$ cursor 
at time step $n$, and $State(n)$ is the state of the machine
at time step $n$.  Mathematically, this models all the cells
of the tapes for all time steps.   The functions 
will be defined as follows.
At time $t = 0$, 
$$\align
   {}&\forall i, Cursor_i(0) = 0 , \quad 
   State(0) = s, \tag3.3 \\
  {}& Tape_1^0(0) = \vartriangleright,  \\
  {}& Tape_1^0(j) = x_j, 1\leq j \leq m, \text{ where }
      x = x_1 x_2 \dots x_m \\
  {}& Tape_1^0(j) = \sqcup \text{ otherwise },
\endalign
$$
for $2 \leq i \leq k$,
$$\align
  {}& Tape_i^0(0) = \vartriangleright,  \tag3.4 \\
  {}& Tape_i^0(j) = \sqcup \text{ otherwise }.
\endalign
$$
For time $t \geq 1$,
if $State(t-1) \neq "yes", "no",$ or $h$, then
first apply the transition function to the corresponding 
cells and state of the tapes at time $t - 1$,
and then update the tapes, cursors, and state.  Mathematically,
it is as follows.  Let
$$\align
  {}&\delta(State(t-1), 
   Tape_1^{t-1}(Cursor_1(t-1)), 
   \dots
   Tape_k^{t-1}(Cursor_k(t-1))) = \tag3.5 \\
   {}&(q; \sigma_1, D_1; \dots ; \sigma_k, D_k),
\endalign
$$ 
then for all $i$,
$$\align
   Cursor_i(t) = {}&Cursor_i(t-1) \text{ if } D_i = - \tag3.6 \\
               = {}&Cursor_i(t-1) + 1 \text{ if } D_i = \rightarrow \\
               = {}&Cursor_i(t-1) - 1 \text{ if } D_i = \leftarrow  , \\
\endalign
$$
$$\align
  Tape_i^t(Cursor_i(t-1)) = {}& \sigma_i, \tag3.7 \\
  Tape_i^t(j) = {}& Tape_i^{t-1}(j), \quad\forall j \neq Cursor_i(t-1),
\endalign
$$ and 
$
   State(t) = q.
$
If $State(t - 1) = "yes", "no",$ or $h$, then
$$\align
   Tape_i^t(n) = {}&Tape_i^{t-1}(n), \quad\forall i, n, \tag3.8\\
   Cursor_i(t) = {}& Cursor_i(t-1), \\
   State(t) = {}&State(t-1).
\endalign
$$
Equation 3.8 says if the machine goes into
the $"yes", "no",$ or $h$ state at time $t - 4$,
then nothing changes from then on.
Now define the function 
$MTime:\left\{State\right\}\to \Bbb N\cup\infty$
as follows.  If there is an $n\in\bar\Bbb N$ such that
$State(n) = "yes", "no"$, or $h$, then 
$$
    MTime = \min\left\{n| State(n) = "yes", "no", \text{ or } h\right\},
    \tag3.9
$$
else, $MTime = \infty$.  Define $Mstate$ to be
$"yes", "no", h$ if there exist some $n\in\Bbb N$ 
such that $State(n) = "yes", "no"$, or $h$ respectively.
Otherwise, $Mstate = \nearrow$
Lastly, define $Mout$ to be the content of the
last tape (not including the infinite continual $\sqcup$ strings) 
if the machine at some point enters the 
$"yes", "no",$ or $h$ state and define it to 
be $\emptyset$ otherwise  
(the model does not allow an output if the machine
computes forever). 
Thus, given an input $x$,
$M(x)$ produces
$$\align
 {}&\left\{Tape_1^0, Tape_2^0, \dots ,Tape_k^0\right\},
  \left\{Tape_1^1, Tape_2^1, \dots ,Tape_k^1\right\},
    \dots \tag3.10 \\
 {}&\left\{Cursor_1, Cursor_2, \dots Cursor_k\right\}\\
 {}&State, Mtime, Mout, Mstate.
\endalign
$$
At this point, a comment is in order.  The superstructure
we are working with is defined over $\Bbb R$.  Some of the
symbols that we used are not elements of $R$.   Since
everything is countable, we can map each symbol to an element 
of $\Bbb R$.  Technically, this has to be done in order
to work with the superstructure.

More generally, let 
$$\align
  {}&TAPE = \tag3.11 \\
     {}&\bigg\{
       \left\{
       \left\{Tape_1^0, Tape_2^0, \dots , Tape_k^0 \right\},
       \left\{Tape_1^1, Tape_2^1, \dots , Tape_k^1 \right\},
       \dots\right\} | k\in\Bbb N, \\
       {}& \Sigma\in\Cal P(\Bbb N)
       \bigg\},
\endalign
$$
where 
$\Cal P\left(\Bbb N\right)$ is the set of all 
subsets of $\Bbb N$, and
for all $i,j$, 
$Tape_i^j:\bar\Bbb N\to\Sigma$.
Now let $FUNC$ be the set
of all functions from $\bar\Bbb N$ to elements of 
$\Cal P\left(\Bbb N\right)$, 
then in terms of the superstructures' language, 
the sets $FUNC$ and $TAPE$ can be characterized
by
$$\align
  {}&\left(\forall x \in FUNC\right)
    \left(\exists \Sigma\in \Cal P\left(\Bbb N\right)\right)
     \bigg[\left(|\Sigma| \in \Bbb N\vee
        |\Sigma| = |\Bbb N|\right)\wedge
     \left(x:\bar\Bbb N \to \Sigma\right)\bigg],\tag3.12 \\
  {}&(\forall x\in TAPE)
       \left(\exists \Sigma\in \Cal P\left(\Bbb N\right)\right)
       \left(\exists k\in\Bbb N\right)\\
       {}&\left(\forall i\in\left\{1,\dots , k\right\}\right)
        \left(\forall j\in\bar\Bbb N\right)
         \left(\exists Tape_i^j\in FUNC\right)\\
  {}&\bigg[
       \left(x = \left\{
        \left\{Tape_1^0, Tape_2^0, \dots , Tape_k^0 \right\},
         \left\{Tape_1^1, Tape_2^1, \dots , Tape_k^1 \right\},
         \dots\right\}\right)\wedge \\
  {}&\left(Tape_i^j:\bar\Bbb N \to \Sigma\right)\bigg] ,
\endalign   
$$
where we use the notation $|\Sigma| = |\Bbb N|$
to mean $\Sigma$ is countable infinite.
Similarly, define  
$$\align
   {}&\left(\forall x\in FUNC_{\bar\Bbb N}\right)
     \bigg[x:\bar\Bbb N\to \bar\Bbb N\bigg], \tag3.13\\
  {}&(\forall x\in CURSOR)\left(\exists k\in\Bbb N\right)
        \left(\forall i\in\left\{1,\dots k\right\}\right)
        \left(\exists Cursor_i\in FUNC_{\bar\Bbb N}\right) \\ 
    {}&\bigg[x =
          \left\{Cursor_1, Cursor_2, \dots Cursor_k\right\}
        \bigg], \\
  {}&\left(\forall x\in STATE\right) 
      \left(\exists K\in \Cal P\left(\Bbb N\right)\right)\bigg[
       \left(|K|\in\Bbb N \vee |K| = |\Bbb N|\right)\wedge
       \left(x:\bar\Bbb N:\to K\right)\bigg], \\
  {}&Mtime: \left(\Sigma - \sqcup\right)^*\to 
       \Bbb N\cup\infty,\\
  {}& Mout:
          \left(\Sigma - \sqcup\right)^*\to\left(\Sigma\right)^*
          \cup\emptyset, \\
  {}&Mstate: \left(\Sigma - \sqcup\right)^*
          \to  \left\{"yes", "no", h, \nearrow\right\}. 
\endalign
$$
The reader should note that the sets defined above
are sets that contains "all" functions of the 
particular type.  For sanity of notations,
the descriptions in (3.12) and (3.13) do not
indicate "all".
\proclaim{\bf Definition 3.1 Deterministic Turing Machine}
Given $\delta , K, \Sigma$, and $k$, 
a deterministic Turing machine
is a function
$$\align
 {}&M_{\delta , K, \Sigma, k}: 
 \left(\Sigma - \sqcup\right)^* \to \tag3.14 \\
  {}&TAPE \times 
  CURSOR \times STATE
  \times \Bbb N\cup \infty \times \left(\Sigma\right)^*
          \cup\emptyset \times 
          \left\{"yes", "no", h, \nearrow\right\},  
\endalign
$$
where the sets are defined above.   Denote
the set of all deterministic Turing machines by
$$
   DTM = \left\{M_{\delta, K, \Sigma, k}, \forall
               \delta, K, \Sigma, k\right\}. \tag3.15
$$
We will write
$$\align 
  {}&M_{\delta, K, \Sigma, k}\left(x\right) = \tag3.16 \\
  {}&<MTAPES, MCURSORS, MSTATES, Mtime(x), Mout(x), Mstate(x)>.
\endalign
$$

\endproclaim 

We can now apply nonstandard analysis and obtain
the internal deterministic Turing machines. 
For any finite or $*$-finite set $S$, 
we will use the notation $|S|$ to denote
the cardinality of $S$.
Notice that the following are true.
$$\align
   {}&\left(\forall
   M_{\delta, K, \Sigma, k}\in DTM\right) 
   [k\in\Bbb N],\tag3.17 \\
   {}&\left(\forall
   M_{\delta, K, \Sigma, k}\in DTM\right) 
   \left(\exists \Sigma \in \Cal P\left(\Bbb N\right)\right)
   [\left(\sqcup\in\Sigma\right) \wedge 
     \left(\vartriangleright\in\Sigma\right) \wedge 
      \left(|\Sigma|\in\Bbb N \vee |\Sigma| = |\Bbb N|\right)],\\
   {}&\left(\forall
   M_{\delta, K, \Sigma, k}\in DTM\right) 
   \left(\exists K \in \Cal P\left(\Bbb N\right)\right)
   [\left(s\in K\right) \wedge \left(|K|\in\Bbb N\vee |K| = |\Bbb N|\right)],\\
   {}&\left(\forall
   M_{\delta, K, \Sigma, k}\in DTM\right) 
   \left[\delta: K\times\Sigma^k \to
     (K\cup\left\{h, "yes", "no"\right\})
     \times\left(\Sigma\times\left\{\leftarrow,
      \rightarrow, -\right\}\right)^k\right].
\endalign
$$
\proclaim{\bf Definition 3.2 Internal Turing Machines}
Let ${}^*DTM$ denote the set of internal Turing machines,
then ${}^*DTM$ is characterized by the following.
$$\align
   {}&\left(\forall
   M_{\delta, K, \Sigma, k}\in {}^*DTM\right)
   [k\in\Bbb {}^*N],\tag3.18 \\
   {}&\left(\forall
   M_{\delta, K, \Sigma, k}\in {}^*DTM\right)
   \left(\exists \Sigma \in {}^*\Cal P\left(\Bbb N\right)\right)
   [\left(\sqcup\in\Sigma\right) \wedge
     \left(\vartriangleright\in\Sigma\right) \wedge
      \left(|\Sigma|\in\Bbb {}^*N \vee |\Sigma| = |{}^*\Bbb N|\right)],\\
   {}&\left(\forall
   M_{\delta, K, \Sigma, k}\in {}^*DTM\right)
   \left(\exists K \in {}^*\Cal P\left(\Bbb N\right)\right)
   [\left(s\in K\right) \wedge 
   \left(|K|\in\Bbb {}^*N\vee |K| = |{}^*\Bbb N|\right)],\\
   {}&\left(\forall
   M_{\delta, K, \Sigma, k}\in {}^*DTM\right)\\
   {}&\left[\delta: K\times\Sigma^k \to
     (K\cup\left\{h, "yes", "no"\right\})
     \times\left(\Sigma\times\left\{\leftarrow,
      \rightarrow, -\right\}\right)^k\right], \\
   {}&\left(\forall x \in {}^*FUNC\right)
    \left(\exists \Sigma\in {}^*\Cal P\left(\Bbb N\right)\right)
     \bigg[\left(|\Sigma| \in {}^*\Bbb N\vee 
      |\Sigma| = |{}^*\Bbb N|\right)\wedge
     \left(x:{}^*\bar\Bbb N \to \Sigma\right)\bigg],\\
  {}&(\forall x\in {}^*TAPE)
       \left(\exists \Sigma\in {}^*\Cal P\left(\Bbb N\right)\right)
       \left(\exists k\in {}^*\Bbb N\right)\\
  {}&\left(\forall i\in\left\{1,\dots , k\right\}\right)
        \left(\forall j\in {}^*\bar\Bbb N\right)
         \left(\exists Tape_i^j\in {}^*FUNC\right)\\
  {}&\bigg[
       \left(x = \left\{
        \left\{Tape_1^0, Tape_2^0, \dots , Tape_k^0 \right\},
         \left\{Tape_1^1, Tape_2^1, \dots , Tape_k^1 \right\},
         \dots\right\}\right) \wedge \\
      {}&\left(
          Tape_i^j: {}^*\bar\Bbb N\to\Sigma\right)
       \bigg] , \\
  {}&\left(\forall x\in {}^*FUNC_{\bar\Bbb N}\right)
     \bigg[x:{}^*\bar\Bbb N\to {}^*\bar\Bbb N\bigg],\\
\endalign
$$
$$\align
  {}&(\forall x\in {}^*CURSOR)\left(\exists k\in {}^*\Bbb N\right)
        \left(\forall i\in\left\{1,\dots k\right\}\right)
        \left(\exists Cursor_i\in {}^*FUNC_{\bar\Bbb N}\right) \\
    {}&\bigg[x =
          \left\{Cursor_1, Cursor_2, \dots Cursor_k\right\}
        \bigg], \\
   {}&\left(\forall x\in {}^*STATE\right)
      \left(\exists K\in {}^*\Cal P\left(\Bbb N\right)\right)\bigg[
       \left(|K|\in {}^*\Bbb N \vee |K| = |{}^*\Bbb N|\right)\wedge
       \left(x:{}^*\bar\Bbb N:\to K\right)\bigg], \\
  {}&{}^*Mtime: \left(\Sigma - \sqcup\right)^*\to
       {}^*\Bbb N\cup \infty,\\
  {}& {}^*Mout:
          \left(\Sigma - \sqcup\right)^*\to\left(\Sigma\right)^*
          \cup\emptyset, \\
  {}&{}^*Mstate: \left(\Sigma - \sqcup\right)^*
        \to  \left\{"yes", "no", h, \nearrow\right\}.
\endalign
$$
Finally,
$$\align
 {}&\left(\forall M_{\delta , K, \Sigma, k}\in {}^*DTM\right)
  \bigg[
   M_{\delta , K, \Sigma, k}:
    \left(\Sigma - \sqcup\right)^* \to \tag3.19\\
  {}&{}^*TAPE \times
     {}^*CURSOR \times {}^*STATE
     \times {}^*\Bbb N \cup \infty\times \left(\Sigma\right)^*
          \cup\emptyset \times
          \left\{"yes", "no", h, \nearrow\right\}\bigg].
\endalign
$$
\endproclaim
Basically, the internal Turing machines are allowed
to work with $*$-finite quantities.  This allows the 
internal Turing machine to become infinitely more powerful
than the classical Turing machines.

A language $L \subset \left(\Sigma - \sqcup\right)^*$
is said to be decided by a deterministic Turing
machine $M$ if and only if
$\forall x\in L$, $Mstate(x) = "yes"$.  The complexity
class $P$ is the set of all languages that are decidable
in polynomial time.  In our set theoretic notation,
the definition is as follows.
\proclaim{\bf Definition 3.3 $P$} The complexity 
class $P$ consists of all the languages that can
be decided by deterministic Turing machines in
polynomial time in the length of the input.  In
other-words,
$$\align
  {}&\left(\forall L\in P\right)
  \left(\forall x\in L\right)
  \left(\exists p\in \bar\Bbb NPOLY\right)
  \left(\exists M_{\delta , K, \Sigma, k}^L \in
   DTM\right) \tag3.20 \\
   {}&\bigg[Mstate(x) = "yes" \wedge Mtime(x) 
     \leq p\left(|x|\right)
   \bigg], \\
  {}&\left(\forall L\in P\right)
  \left(\exists p\in \bar\Bbb NPOLY\right)
  \left(\exists M_{\delta , K, \Sigma, k}^L \in
   DTM\right) 
    \left(\forall y\in \left(\Sigma - \sqcup\right)^*\right)\\
   {}&\bigg[\left(Mstate(y) = "yes" \wedge Mtime(y) 
     \leq p\left(|y|\right)\right) \to y\in L
   \bigg], \\
\endalign
$$
where the Turing machine 
$M_{\delta , K, \Sigma, k}^L$
depends on $L$ and $|x|$ denotes
the length of the input string $x$.
\endproclaim

The internal class ${}^*P$ allows internal Turing
machines to decide internal languages in $*$-polynomial
run time.
\proclaim{\bf Definition 3.4 ${}^*P$} The internal complexity
class ${}^*P$ consists of all the internal languages that can
be decided by internal deterministic Turing machines in
$*$-polynomial time in the length of the input.  In
other-words,
$$\align
  {}&\left(\forall L\in {}^*P\right)
  \left(\forall x\in L\right)
  \left(\exists p\in {}^*\bar\Bbb NPOLY\right)
  \left(\exists M_{\delta , K, \Sigma, k}^L \in
   {}^*DTM\right\} \tag3.21 \\
   {}&\bigg[Mstate(x) = "yes" \wedge Mtime(x)
     \leq p\left(|x|\right)
   \bigg], \\
  {}&\left(\forall L\in {}^*P\right)
  \left(\exists p\in {}^*\bar\Bbb NPOLY\right)
  \left(\exists M_{\delta , K, \Sigma, k}^L \in
   {}^*DTM\right) 
    \left(\forall y\in \left(\Sigma - \sqcup\right)^*\right)\\
   {}&\bigg[\left(Mstate(y) = "yes" \wedge Mtime(y) 
     \leq p\left(|y|\right)\right) \to y\in L
   \bigg], \\
\endalign
$$
\endproclaim

We now show that whatever the classical Turing machines
can do, the internal Turing machines can do in
$*$-polynomial time.
\proclaim{\bf Theorem 3.1} Let $M\in DTM$ be a classical
deterministic Turing machine
and let $L$ be a language for $M$ such that for all $x\in L$,
$Mout(x)\neq \nearrow$.  In
other-words, $M$ eventually halts on input $x$.  
Let ${}^\sigma L$ be the $*$-embedding of $L$, 
$$
   {}^\sigma L = \left\{{}^*x| x \in L\right\}. \tag3.22
$$
Then, on input $x\in {}^\sigma L$, the internal Turing machine ${}^*M$ 
will halt in $*$-polynomial time.  Further,
the internal Turing machine will halt in the same
state as $M$ and output the same output as $M$.
\endproclaim
\demo{proof} Let $\omega \in {}^*\Bbb N$
be an infinite integer, then the polynomial
$\omega n: {}^*\bar\Bbb N\to
{}^*\bar\Bbb N$ is an internal polynomial.
Let $Mtime(x)$ be the runtime of $M$ on input $x\in L$,
this number is standard finite (an element of $\Bbb N$)
since $M$ halts on input $x$.  

Since $M$ halts on $x$,
${}^*M\left({}^*x\right)$ will halt
and the time for
it to halt is $Mtime(x) < \omega |x|$. Further,
on input ${}^*x$, 
${}^*M$ will halt in the same
state as $M$ and output the same output as $M$. \qed
\enddemo
\remark{Remark} In other words, as far as 
${}^*M$ is concerned, on input from $L$ (or
equivalently, the $*$-embedding of $L$), 
${}^*M$ will halt in $*$-polynomial time
regardless of the time for $M$ to halt on
input from $L$.  This says that whatever a
classical deterministic Turing machine does and halts,
there is an internal deterministic Turing machine
that can do the same thing in $*$-polynomial time.   
We have to be very careful in using and interpreting
nonstandard analysis.  This does not mean that
we can use the $*$-transform principle and
conclude that the classical deterministic Turing machines can
do anything in polynomial time since the set
${}^\sigma L$ is an external set.
\endremark

One way to interpret theorem 3.1 is as follows.
We can think of the internal Turing machines as
digital machines that can work with an infinite
number of bits and can compute infinitesimally
close to continuous variables.  For example,
let $r\in [0,1]$
and let 
$$
  r = \lim_{k\to\infty} \sum_{i=0}^k a_i 2^{-i}
   = \lim_{k\to\infty} s_k , \tag3.23
$$
be its base 2 expansion.  Let $\omega\in {}^*\Bbb N$
be a nonstandard infinite integer and $s_{\omega}
\in {}^*\left\{s_k\right\}$, then 
$s_\omega$ is represented by $\omega$ number
of bits and it is infinitesimally close to $r$.
Suppose it takes the internal Turing machine
${}^*M$ $k$ units of time to write $\omega$
number of bits. 
Theorem 3.1 says that on input $x\in {}^\sigma L$, 
${}^*M(x)$ will halt in less than $k|x|$ units of time.
We can think of this as exploiting the internal
Turing machines' ability to efficiently compute
quantities that are infinitesimally close to real
variables.
For more thoughts on time issues, see section 5.

Now we show that the internal deterministic 
Turing machines can
do things which the 
classical deterministic Turing machines can not do.  
In particular,
the internal deterministic Turing machines can decide 
${}^\sigma Halt$, the $*$-embedding of the classical
$Halt$ language.  It is well known that 
$Halt$ can not be decided by the classical deterministic Turing
machines.  As with the previous theorem, we must
be careful how we interpret this because
${}^\sigma Halt$ is an external set.  This does not 
imply that the internal deterministic Turing machines can
decide ${}^*Halt$ because this would imply that
the classical deterministic Turing machine can decide $Halt$
by the transfer principle.  In fact, ${}^*Halt$ can
not be decided by the internal deterministic Turing machines.
Thus, the internal Turing machines are infinitely more
powerful than the classical Turing machines but at the same
time, the internal machines have the same type of limitations
as the classical machines.

\proclaim{\bf Proposition 3.2 Universal Turing machines}
There exists a universal deterministic Turing machine which can simulate
any other deterministic Turing machine in polynomial time.  
In other words,
$$\align
  {}&\left(\exists U_{\delta , \Bbb N, \Bbb N, k}\in DTM\right)
  \left(\forall M_{\delta^{'} , K, \Sigma, k^{'}}
   \in DTM\right)
   \left(\forall x \in \left(\Sigma - \sqcup\right)^*\right)
   \left(\exists p\in \bar\Bbb NPOLY\right) \tag3.24 \\
   {}&\bigg[Uout(M_{\delta^{'} , K, \Sigma, k^{'}};x) = 
      Mout(x) \wedge 
    Ustate(M_{\delta^{'} , K, \Sigma, k^{'}};x) = Mstate(x)
        \wedge \\
     {}& \left(Mtime(x) \neq \infty \to 
           Utime(M_{\delta^{'} , K, \Sigma, k^{'}};x) 
        \leq p\left(Mtime(x)\right)\right)
        \wedge \\
     {}& \left(Mtime(x) = \infty \to 
           Utime(M_{\delta^{'} , K, \Sigma, k^{'}};x) = \infty
        \right)\bigg].
\endalign
$$
\endproclaim
\demo{proof} See [8].\qed
\enddemo

\remark{\bf Remark 3.1} The notation
$U\left(M;x\right)$ implies an encoding
of the deterministic Turing machine $M$ and
input $x$ (for $M$)
as input for the universal Turing machine $U$
Further, the number of symbols and the number
of states for $U$ is allow to be countable
infinite to compensate for arbitrary 
$M_{\delta^{'} , K, \Sigma, k^{'}}$
\endremark

For the rest of this section, we will fix a universal
Turing machine $U$.
\proclaim{\bf Definition 3.5 Halting}
Let $M$ be a deterministic Turing machine
and $x$ be an input for $M$.
Define the language $Halt$ over the
alphabet of $U$, the universal Turing machine as
$$
   \left(\forall <M;x>\in Halt\right)
   \bigg[Mstate(x) \neq \nearrow\bigg]\tag3.25 
$$
In otherwords, $Halt$ consists of the
encoding of Turing machines $M$ and input $x$ for
the universal Turing machine $U$ such
that $M$ eventually halts on input $x$.
\endproclaim

We now modify the universal Turing machine $U$
into another Turing machine.  Let $n\in \Bbb N$,
define the Turing machine $U_n$ as follows.
On input $M;x$, $U_n$ simulates $M$ with input $x$.
If $U_n(M;x)$ halts in time less than $n$, then
$U_n$ computes as $U$, otherwise, $U_n(M;x)$
goes into the $"no"$ state at step $n$.  
In terms of our set theoretic notation, this can
be written as (with an abuse of notation)
$$\align
   {}&\left(\forall n\in N\right)
   [U_n \in DTM]; \tag3.26 \\
   {}&\left(\forall M_{\delta, K, \Sigma, k} \in DTM\right)
      \left(\forall x \in \left(\Sigma - \sqcup\right)^*\right) \\
   {}&
     [\left(Mstate(x) = "yes", "no", h \wedge 
       Mtime(x) < n\right)
          \to \\ 
    {}&\left(U_nstate(M;x) = "yes" 
            \wedge U_ntime(M;x) < n\right)
    \wedge \\
    {}&\left(else \to U_nstate(M;x) = "no" \wedge 
               U_ntime(M;x) = n\right) ] .
\endalign
$$

\proclaim{\bf Theorem 3.3} Let 
$$
  {}^\sigma Halt =
   \left\{<{}^*M; {}^*x> | <M; x> \in Halt\right\} , \tag3.27
$$
be the $*$-embedding of $Halt$, then there is an internal
deterministic Turing machine $U_\omega$ which on input
$<{}^*M, {}^*x> \in {}^\sigma Halt$ will halt in the state
$"yes"$.  Further, if $<{}^*M, {}^*x> \notin {}^\sigma Halt$,
$U_\omega$ will halt in the $"no"$ state.
\endproclaim
\demo{Proof} The $*$-transform of (3.26) is 
$$\align
   {}&\left(\forall n\in {}^*N\right)
   [U_n \in {}^*DTM]; \tag3.28 \\
   {}&\left(\forall M_{\delta, K, \Sigma, k} \in {}^*DTM\right)
      \left(\forall x \in \left(\Sigma - \sqcup\right)^*\right) \\
   {}&
     [\left(Mstate(x) = "yes", "no", h \wedge
       Mtime(x) < n\right)
          \to \\
    {}&\left(U_nstate(M;x) = "yes"
            \wedge U_ntime(M;x) \leq n\right)
    \wedge \\
    {}&\left(else \to U_nstate(M;x) = "no" \wedge
               U_ntime(M;x) = n\right) ] .
\endalign
$$
Now let $n = \omega \in {}^* \Bbb N$ be a nonstandard infinite
integer, then $U_\omega$ is an internal Turing machine.
Suppose $<M;x> \in Halt$, 
${}^*M({}^*x)$ produces ${}^*Mstate(x) = "yes"$, $"no"$,
or $h$, and ${}^*Mtime(x) = t$ for some standard finite $t$.
Since $t$ is standard finite, we have $t < \omega$, which
implies that $U_\omega(M;x)$ halts with a $"yes"$ at
time $t$.  
Further, if $M\in DTM$ and $M(x)$ does not halt, 
then ${}^*M({}^*x)$ will not halt
and $U_\omega(M;x)$ will halt with a $"no"$ at time
$\omega$.  \qed
\enddemo

\proclaim{\bf Theorem 3.4} The language $Halt$ is
not decidable by the classical deterministic Turing
machines $DTM$.
\endproclaim
\demo{Proof} See [8] \qed
\enddemo

\proclaim{\bf Theorem 3.5} The internal set ${}^*Halt$ is
not decidable by the internal deterministic Turing
machines ${}^*DTM$.
\endproclaim
\demo{Proof} This is a property of the $*$-transform \qed
\enddemo

\subhead{\bf 4. Real computations and Asymptotic Behaviors}
\endsubhead  The internal Turing machines could be useful 
in determining asymptotic behaviors of real number
computations.  We will use the real number computation
model proposed in [6].  The classical Turing machines
are modeled with discrete mathematics.  It would be
unnatural to use limits to obtain asymptotic behaviors.
The internal Turing machine model is quite natural
for this purpose since it keeps all the 
combinatoric structures
of the classical model.  This could be a bridge for
the gap between the discrete model of computational
complexity theory and the continuous variable model
of physical theories.  We will illustrate the latter idea
in the next section.

We now set the notations and foundations for the
real number computation model proposed in [6].
A dyadic rational number $d$ is a number of the
form $d = \frac{m}{2^n}$, where $m\in\Bbb Z$ and
$n \in \Bbb N \cup 0$.  Denote by 
$$
   D_n = \left\{m*2^{-n} | m \in \Bbb Z\right\}, \tag4.1
$$
the set of dyadic rational numbers with
precision $n$ and $D = \cup_{n = 1}^\infty D_n$ the
set of all dyadic rational numbers.  
\proclaim{\bf Definition 4.1 Classical Computable Real Numbers}
Let $x\in\Bbb R$, $x$ is said to be computable in polynomial
time if the
following is true.  There exists a function
$\Phi:\Bbb N \to D$ such that for all
$n\in\Bbb N$,
the precision $prec\left(\Phi(n)\right) = n$, 
$|\Phi(n) - x| \leq 2^{-n}$, and there exists
a Turing machine that computes $\Phi(n)$ in polynomial time.
In other-words,
$$\align
  {}&(\forall n\in \bar\Bbb N)
  [\Phi(n) \in D_n],\tag4.2 \\
  {}&(\forall n\in \bar\Bbb N)
  [|\Phi(n) - x| \leq 2^{-n}],\\
  {}&\left(\exists M \in DTM\right)
      \left(\exists p\in\bar\Bbb NPOLY\right)
       \left(\forall n\in \bar\Bbb N\right) 
   [Mout(n) = \Phi(n) \wedge Mtime(n) \leq p(n)] .
\endalign
$$
\endproclaim
\proclaim{\bf Theorem 4.2} Let $x\in\Bbb R$ be 
computable in polynomial time in the sense of 
definition 4.1, then there exists an internal
Turing machine which can compute an internal
dyadic number 
$d \in {}^*D$ that is infinitesimally
close to $x$.  Further, this computation is done in
$*$-polynomial time.
\endproclaim
\demo{Proof} Applying the $*$-transform
to (4.2) yields
$$\align
   {}&(\forall n\in {}^*\bar\Bbb N)
  [{}^*\Phi(n) \in {}^*D_n], \tag4.3 \\
  {}&(\forall n\in {}^*\bar\Bbb N)
  [|{}^*\Phi(n) - x| \leq 2^{-n}],\\
  {}&\left(\exists M \in {}^*DTM\right)
      \left(\exists p\in {}^*\bar\Bbb NPOLY\right)
       \left(\forall n\in {}^*\bar\Bbb N\right) \\
   {}&[Mout(n) = {}^*\Phi(n) \wedge Mtime(n) \leq p(n)] .
\endalign
$$
In particular, let $n = \omega$ be an infinite
integer, then $|{}^*\Phi\left(\omega\right) - x| 
  \leq 2^{-\omega}$ and 
$2^{-\omega}$ is an infinitesimal.  \qed
\enddemo

For any
real number $x\in\Bbb R$,
the internal Turing machines might be able to output
a $*$-finite number of bits which could
represent an element (in ${}^*\Bbb R$) that
is infinitesimally close to $x$.
However, unless $x$ has a standard finite
numerical representation, to output ${}^*x$ (the 
$*$-embedding of $x$) would require
the machine to write a "${}^*\infty$"
number of bits on one of its tapes.  This is the
same limitation that the classical Turing machines
have.  Thus, for the internal Turing machines,
in general it is not possible to compute the standard part
function since $st(x)$ could requires an infinite amount of
time to output.  This is as expected since the standard part function
is an external function.  As far as outputting the exact value of 
${}^*x$ is concerned, the best we could hope for
is that an observer looks at the output of the 
internal machine and then apply the standard part
operation to the machine's output.
As far as humans are concerned, assuming that
we can only measure a standard finite number of bits,
we will never be able to observe the exact value of $x$.
If the internal Turing machines were a reasonable model
for nature's computation structure, then the fact that
it can never output the exact value of $x$ indicates that
it might be more suitable to model physical phenomenons
with $*$-finite nonstandard analysis rather than
continuous variables in $\Bbb R$.

\subhead{\bf 5. Probabilistic Turing Machines and 
Simulating Finite State Quantum Mechanics}
\endsubhead  In this section, we take on the idea
that the internal Turing machine model is a reasonable
model for nature's behind the scene physical computations.  
We assume that the working tapes of the internal
Turing machines are hidden from us but we can
observe the machines' output tapes when we perform
a measurement. 
We will use an internal probabilistic Turing machine
to simulate time independent finite state quantum mechanics in
$*$-polynomial time.  
As far as we know, there is no evidence that 
nature uses internal Turing machines to 
compute its physical processes. 
Hence, we will think of this section
as a thought experiment.   

We believe that there
are a few reasons why it is interesting
to do this thought experiment.  The
first is that the internal Turing machines
are extensions of the classical ones,
and the internal ones are capable of doing
computations that are infinitesimally
close to continuous variables. 
This could be a digital bridge between the
Church-Turing thesis, experimental science,
and continuous variable modeling.  The second
is that the algorithm described below (using
internal probabilistic Turing machine model) is not
at odds with instantaneous collapse of 
the wave function and 
no information
can travel faster than the speed of light.
Further, it suggests that the classical
definition of Turing machines could be much
more fundamental than previously thought;
the definition of the classical Turing machines
might be more capable of dealing with quantum
phenomenons than previously thought.
Recently, there has been
much research activity in quantum computing and
quantum Turing machines (see [1],[2], [10], [11], 
and references within).  
It is now widely
believed that the quantum Turing machines are more
powerful than the probabilistic Turing machines.
The ideas in section could be of interest for quantum
computing research.

Classically, a probabilistic Turing machines
consists of two elements $<M, \Psi>$,
where $M$ is a deterministic Turing machine
and $\Psi$ is a random coin flip.  The operation
of the machine is roughly described as follows
(see [4] for a full description).
At each step of the computation, $\Psi$ flips
its coin and with probability $1/2$ outputs a 0
or 1 on a special random bit tape, then the machine
$M$ makes its next move according to all its
tapes including the random bit tape.  Let
$PTM$ denote the set of all probabilistic
Turing machines, without spelling out
the superstructure set theoretic definition of
the probabilistic Turing machines, we can characterize
$PTM$ as follows.
$$\align
 {}&(\forall <M, \Psi> \in PTM)
 [M\in DTM \wedge prob(\Psi = 1) = .5 \wedge prob(\Psi = 0) = .5] .
  \tag5.1
\endalign
$$ 
Its $*$-transform is
$$\align
 {}&(\forall <M,\Psi> \in {}^*PTM)
 [M\in {}^*DTM \wedge prob(\Psi = 1) = .5 \wedge prob(\Psi = 0) = .5] .
  \tag5.2
\endalign
$$ 
Thus, 
\proclaim{\bf Definition 5.1 Internal Probabilistic
Turing Machines} An internal probabilistic Turing machines consists of
two elements $<M,\Psi>$ where $M\in {}^*DTM$ is an internal
deterministic Turing machine and 
$\Psi$ is a $0,1$ coin flip with probability
$1/2$.  The machine operates as follows.  At each step,
$\Psi$ flips a coin and outputs a 0 or 1 on its random bit tape,
then the machine $M$ makes its next move according to all its tapes
including the random bit tape.
\endproclaim

We now proceed to simulate finite state quantum mechanics.
This will be done with two algorithms, the evolution
and measurement algorithm.  We first describe the evolution.
Let $U = U_R + i U_I$ be an $n$ by $n$ time independent unitary matrix 
where $U_R$ and $U_I$ are the real and imaginary parts
of $U$.  Let
$v^{in} = \sum_{k = 1}^n a^k e_k$ be a quantum state
where the $e_k$'s form a basis for the underlying Hilbert space.
Let $\omega \in {}^*\Bbb N$ be an infinite nonstandard
integer.  The algorithm is independent of which
infinite nonstandard integer is picked. 
For $1 \leq k,j\leq n$, let
$\left(U_{R\omega}\right)_{k,j},
 \left(U_{I\omega}\right)_{k,j}
  \in D_\omega$ 
   be the approximation of  
$\left(U_R\right)_{k,j}$ and 
$\left(U_I\right)_{k,j}$  
by elements of $D_\omega$ (the internal set of 
dyadic rational numbers with precision $\omega$
as defined in the previous section) such that 
the approximation
is infinitesimally close.  
Denote this by
$$
  U_R \approx U_{R\omega}, U_I \approx U_{I\omega} ,
  \tag5.3
$$
where $U_{R\omega}$ and $U_{I\omega}$
denote the matrices of the corresponding $\omega$ precision
approximations.  Similarly, let $v_\omega^{in}$
be an $\omega$ precision approximation to
$v^{in}$ and denote it by $v^{in} \approx v_\omega^{in}$.
Finally, let
$U_\omega = U_{R\omega} + i U_{I\omega}$,
and 
$$
  Uv^{in} = v^{out} = \sum_{k = 1}^n b^k e_k
  \tag5.4
$$

We now describe the input to the internal 
probabilistic Turing machine which will 
simulate the evolution of 
finite state quantum mechanics.  
Recall that the internal Turing machines
are allowed to have a $*$-finite number of 
tapes, this allows for enough memory to deal
with $n$ number of quantum states for arbitrary
standard finite $n$.
We will take $U_\omega$,  $v_\omega^{in}$, and
$1^\omega$ (a string consisting of an $\omega$ number of 1's) as
the input to the machine.  This input could be 
computed in $*$-polynomial time by another internal machine
as described in the previous section or it could be obtained
from an internal oracle machine.  
We might think of 
$v^{in}$ as the initial state of a quantum experiment,
$U$ as the evolution operator corresponding to
the experiment, the environment dictates $1^\omega$
(or $\omega$ is a universal constant, or $\omega$ is
an intrinsic property of the probabilistic Turing
machine),
and the internal probabilistic Turing machine
reads in $v^{in}$, $U$ as $v_\omega^{in}$,
$U_\omega$ and then computes.   In any case,
we will assume that the machine
obtains $U_\omega$, $v_\omega^{in}$, and $1^\omega$ as input.
This assumption is justified by the thesis
that nature takes an initial state and the environment
as input and computes the physical evolution.
The size of the above input is linear $\omega$.

Upon receiving the input, the internal machine computes as follows.
First, it computes $\bar\omega = \lfloor\log_2\omega\rfloor$,
which is an infinite nonstandard integer.  This computation
can be done in $*$-polynomial time in $\omega$ since its classical
equivalent can be done in polynomial time, i.e.,
$$
  (\exists M \in DTM)
  (\exists p \in \bar\Bbb NPOLY)
  (\forall k\in\Bbb N)
  [Mout(k) = \lfloor\log_2 k\rfloor \wedge
   Mtime(k) \leq p(k)] .
  \tag5.5
$$
Next, the internal machine computes
$U_\omega v^{in}_\omega$.   
Let us denote
the output of this stage by
$\sum_{k = 1}^n b^k_{\omega} e_k$.  
This computation can also be computed in
$*$-polynomial time in $\omega$ since its classical
equivalent can be computed in polynomial time,
i.e., 
$$\align
  {}&(\exists M \in DTM)
  (\exists p \in \bar\Bbb NPOLY)
     (\forall k\in\Bbb N) \tag5.6 \\
  {}&[Mout(U_k, v^{in}_k) = U_kv^{in}_k \wedge
    Mtime(U_k, v^{in}_k) \leq p\left(k\right)] .
\endalign
$$

If the quantum experiment or environment
does not perform a measurement, the algorithm
writes the out state computed above to the
output tape and halts, otherwise, the algorithm
proceeds to the second part, the measurement algorithm.  
Notice that if it halts, 
then the output state vector will be entry-wise
infinitesimally close to $v^{out}$, and we can
interpret this as nature putting the quantum 
system into a state that is infinitesimally close
to $v^{out}$.
Now suppose a measurement is performed.
After computing the out state, the machine then computes
$|b_{\omega}^k|^2, 1\leq k \leq n$ with $\bar\omega$
precision.  This too can be done in $*$-polynomial
time since its classical equivalent can be done in
polynomial time.  Let us denote the output of this
stage by $pr_k, 1 \leq k \leq n$.  Notice that 
$$
   pr_k \approx |b^k|^2,
   |b^k|^2 \leq 1, \sum_{k = 1}^n pr_k \approx
   \sum_{k = 1}^n |b^k|^2 = 1 ,\tag5.7
$$ since $\bar\omega$ is nonstandard
infinite (the error in the $\bar\omega$ roundoff
is less than $2^{-\bar\omega}$).  
Further, $pr_k$ having precision $\bar\omega$
means that it is of the form $\dfrac{m_k}{2^{\bar\omega}}$
for $m_k\in {}^*\bar\Bbb N$, or more generally, 
$pr_k \in {}^*D_{\bar\omega}$
where ${}^*D_{\bar\omega}$ is defined in the previous 
section.

The machine now uses its coin flip ability and
performs the last stage of the computation.  
If there is a $k$ such that $1 \leq pr_k$,
then the machine outputs $k$ and halts.  This
can be done in $*$-polynomial time since its classical
equivalent can be done in polynomial time. 
If for all $k$, $pr_k < 1$, then the machine proceeds
as follows.  Recall that $pr_k = \frac{m_k}{2^{\bar\omega}}$,
and $\sum_{k = 1} pr_k \approx 1$.  Let
$\sum_{k = 1} pr_k = 1 + \epsilon$, where
$\epsilon$ is an infinitesimal.  The machine
computes the internal number $2^{\bar\omega}$,
then computes $m_k$ for all $k$, and then 
computes $\sum_{k = 1}^n m_k = T$. This can
be done in $*$-polynomial time in $\omega$ 
since $\bar\omega = \lfloor\log_2\omega\rfloor$.  The machine
now flips its coin $T$ number of times, outputs
a state $k$ according to the probability distribution
$\left\{\frac{m_j}{T}\right\}$, and then halts.  The last
step can be done in $*$-polynomial time in 
its input size
since 
$$
   T = \sum_{k = 1}^n m_k = 2^{\bar\omega}\sum_{k = 1}^n pr_k
   < 22^{\bar\omega} \leq 2\omega . \tag5.8
$$
We now need to show that the probability distribution
$\left\{\frac{m_j}{T}\right\}$ is infinitesimally close 
to the distribution $\left\{\frac{m_k}{2^{\bar\omega}}\right\}$.
This is true because
$$
   \frac{m_k}{\sum_{j} m_j} =
   \dfrac{\frac{m_k}{2^{\bar\omega}}  }
    {\frac{\sum_{j} m_j}{2^{\bar\omega}}} =
   \dfrac{\frac{m_k}{2^{\bar\omega}}  }
    {1 + \epsilon} ,
    \tag5.9
$$   
which implies
$$
   \frac{m_k}{\sum_{j} m_j} \approx
   \frac{m_k}{\sum_{j} m_j} \left(1 + \epsilon\right) =
   \frac{m_k}{2^{\bar\omega}} .
   \tag5.10
$$  
Equations (5.7) and (5.10) imply that for all $k$,
$\frac{m_k}{T}$ is infinitesimally close to
$|b^k|^2$.  Finally, since each stage of the computation
can be done in $*$-polynomial time, the complete
algorithm halts in $*$-polynomial time.  Notice that
if a partial measurement is performed, then the Turing
machine must compute the output state after the partial
measurement.  For our purpose, we will not need to do this.

The result of the above algorithm is that it outputs
a state $k$ with probability infinitesimally
close to $|b^k|^2$, which is the probability
dictated by the theory of quantum mechanics. 
Further, the algorithm halts in $*$-polynomial time.
As an observer, a human can look at the output
tape of the internal Turing machine and measures
a standard finite state, namely a state $e_k$
where $1 \leq k \leq n$.  Further, as an observer,
a human can only perform the experiment 
or run the above algorithm a standard
finite number of times.  This would imply that
the observer will never be able to detect the
fact that the statistics of the output of
the above algorithm is only infinitesimally close
to the result dictated by theory of 
quantum mechanics.

There are a few interesting things that we
can conclude by taking on the idea that 
nature behaves this way.
The first is that given an input, 
nature requires time to compute the output in
both the evolution and measurement algorithm.  In the
most loose interpretation, this might be related to
no information can travel faster than speed of light.
In which case, the time required to perform
one step of the above computation is related to the speed of light.
On the other hand, collapse of the wave function says otherwise.
It says that measurements on spatially separated
quantum systems can instantaneously influence one another.
At first thought, the above algorithm is at odds
with collapse of the wave function, but in fact, that need
not be the case.  This is because the definition of 
a multi-tape deterministic Turing machine allows at each time
step the simultaneous
reading and writing of one cell on each of its tapes.
For example, suppose Alice and Bob each has a qubit.
They run the the above evolution algorithm with the 
appropriate unitary operator and obtain
a state that is infinitesimally close to a Bell
state, i.e., the internal Turing machine computes
$$
  b_{\omega}^1|0\rangle|0\rangle +
  b_{\omega}^2|1\rangle|1\rangle \approx
  \frac{1}{\sqrt{2}}|0\rangle|0\rangle +
  \frac{1}{\sqrt{2}}|1\rangle|1\rangle ,
  \tag5.11
$$
and writes the state
$b_{\omega}^1|0\rangle|0\rangle +
  b_{\omega}^2|1\rangle|1\rangle$ on a working  
tape (or puts the qubits into the Bell state) and 
then temporarily halts until
further notice to perform the measurement
algorithm.  At this point, we can think of nature
putting Alice and Bob's qubits into the Bell state,
and we assume that 
the machine has two output tapes, one for
each qubit. 
Alice and Bob now each takes 
their output tape (their qubits)
and they separate light years apart.  
After the spatial separation, Bob (or Alice) gives the 
internal Turing machine the go to perform the
measurement stage of the computation.   Suppose
the machine uses it coin flip ability and
outputs $|0\rangle|0\rangle$.  At output,
the machine writes the state $|0\rangle$ on
Bob's tape and "simultaneously" writes
$|0\rangle$ on Alice's tape.   The definition
of Turing machine does not prevent this from
happening even though the tapes are separated
light years apart.  Hence, the Turing machine
model has the ability to model instantaneous transmission
of information during the read and write operation
at each time step.  This suggests that the Turing machine
model might be more fundamental than previously thought.

\comment
Let us look at the above spatial separation
from an classical physics point of view.
Suppose we physically build a deterministic Turing machine
using classical physics to efficiently compute
a desired output upon an input.  Further, suppose the machine
consists of two working tapes and one output tape 
and the tapes are reasonably separated and 
wirelessly communicate
with each.  Now
we take the tapes and separate them 
light years apart from each other and
from the machine itself.
From a practical point of view, no one would argue that
after the separation, we can continue to efficiently
compute with the machine but the Turing machine model
does not distinguish the difference between the 
pre-separated and post-separated machines.  This
suggests that in this extreme case,
the Turing machine model might be
more suitable for quantum physics than classical physics,
and this would imply that the Turing machine model is
more fundamental than previously thought since
quantum mechanics is more fundamental than classical
mechanics.  
\endcomment

As a final note for this section, 
notice that the standard part
operation is not needed since the output
of the algorithm is standard finite.  
Further, the infinite
number of bits of computations 
are completely oblivious to the
the human observer since the working tapes
of the Turing machines are hidden from
the observer.   
Thus, the infinite
number of bits of computation is a black
box for the observer.  
Finally, the working tapes are hidden but 
no hidden variables are introduced in this thought
experiment.  In other words, at no time before the measurement is
performed (more precisely, the coin flips) 
there exist variables which if know will completely
determine the outcome of the output state.

\subhead{\bf 6. Nondeterministic Turing Machines}
\endsubhead
In this last section, we show that if $P \neq NP$,
then there exists problems which the internal
Turing machines can solve but not in $*$-polynomial
time.  This is basically a property of the
$*$-transform.  In the previous sections,
we mainly dealt with problems that
the internal Turing
machines can solve in $*$-polynomial time 
(except deciding ${}^*Halt$, which can not
be decided).   Thus,
the internal Turing machines are very powerful
but they also have limitations similar to
the classical Turing machines.  In terms of the
physical, if the internal Turing machines properly
model nature's computational power and if nature
views polynomial time in a sense similar to ours,
then nature would favor physical processes that
are computable in $*$-polynomial time.

The nondeterministic Turing machines are similar to the
deterministic ones except that the transition function
$\delta$ is allowed to be a transition relation.  
For notation convenience, we will just sketch
the set theoretic definition and then apply the
$*$-transform and obtain the internal nondeterministic
Turing machines.  For a $k$ tape nondeterministic Turing machine,
we have 
$$ 
  \delta \subset K\times\Sigma^k \times
     (K\cup\left\{h, "yes", "no"\right\})
     \times\left(\Sigma\times\left\{\leftarrow,
      \rightarrow, -\right\}\right)^k .
   \tag6.1
$$
For example, at time $t = 0$, the tape configuration would
be given by
$$
   \left\{Tape_1^0, Tape_2^0, \dots , Tape_k^0 \right\}.
   \tag6.2
$$
The transition relation will then take the Turing
machine into a computational tree.  If the relation
takes the machine into three configurations, then at time
$t = 1$, we would get three configurations
$$\align
   \bigg\{
   {}&\left\{Tape_1^{1,1}, Tape_2^{1,1}, \dots , Tape_k^{1,1} \right\}, 
       \tag6.2 \\
   {}&\left\{Tape_1^{1,2}, Tape_2^{1,2}, \dots , Tape_k^{1,2} \right\}, \\
   {}&\left\{Tape_1^{1,3}, Tape_2^{1,3}, \dots , Tape_k^{1,3} \right\}
    \bigg\},
\endalign
$$
and similarly for the cursors, states, etc.  At each time step,
the machine would branch off from the configurations of the previous time
step and continues with the computational tree. 
The machine halts if one
of the configurations (computational branches) halts, 
otherwise, it computes forever.  
If the machine halts and one of the halting branches 
halts in $"yes"$ state, then the machine is said to
accept the input $x$.  In which case, we will
write $Mstate(x) = "yes"$.  Notice that there could
be different halting states for different computational
paths.  The time of the computation is somewhat vague
since there are many computational paths where some
of which might halt or some of which might compute forever.
To make the definition of computational time 
complete, we define three computational times, the time
for it to halt into the $"yes", "no"$ and $h$ state.
They are defined as the minimum amount of time over all
configurations that goes into the $"yes", "no"$ and
$h$ state respectively; if the machine never goes into
any of three states, we will define the time to be
$\infty$ for that particular state.  We will denote them
by $Mtime_y(x)$, $Mtime_n(x)$ and $Mtime_h(x)$.
The output of the machine also has this ambiguity.
One way of defining the output is to restrict
the output to $y$ if all computational branches
that halts in the $"yes"$ state outputs $y$ (see [4]).  For our
purpose, we will not be needing the output of the machine.
We will leave the definition of the output open.  
We will mainly be interested in the machine
going into the $"yes"$ state.  

The set of all nondeterministic Turing machines
can be cast into set theoretic languages as
in the previous section for deterministic
Turing machines.  For notation sanity,
we will not spell this out.
The set of all nondeterministic Turing machines
will be denoted by $NDTM$.  Its $*$-transform,
the set of internal nondeterministic
Turing machines will be denoted ${}^*NDTM$.  
\proclaim{\bf Definition 6.1 $NP$} The complexity
class $NP$ consists of all the languages that can
be decided by nondeterministic Turing machines in
polynomial time in the length of the input.  In
other-words,
$$\align
  {}&\left(\forall L\in NP\right)
  \left(\forall x\in L\right)
  \left(\exists p\in \bar\Bbb NPOLY\right)
  \left(\exists M_{\delta , K, \Sigma, k}^L \in
   NDTM\right) \tag6.3 \\
   {}&\bigg[Mstate(x) = "yes" \wedge Mtime_y(x)
     \leq p\left(|x|\right)
   \bigg], \\
  {}&\left(\forall L\in NP\right)
  \left(\exists p\in \bar\Bbb NPOLY\right)
  \left(\exists M_{\delta , K, \Sigma, k}^L \in
   DTM\right)
    \left(\forall y\in \left(\Sigma - \sqcup\right)^*\right)\\
   {}&\bigg[\left(Mstate(y) = "yes" \wedge Mtime_y(y)
     \leq p\left(|y|\right)\right) \to y\in L
   \bigg]. \\
\endalign
$$
\endproclaim

\proclaim{\bf Definition 6.2 ${}^*NP$} The internal complexity
class ${}^*NP$ consists of all the internal languages that can
be decided by internal non-deterministic Turing machines in
$*$-polynomial time in the length of the input.  In
other-words,
$$\align
  {}&\left(\forall L\in {}^*NP\right)
  \left(\forall x\in L\right)
  \left(\exists p\in {}^*\bar\Bbb NPOLY\right)
  \left(\exists M_{\delta , K, \Sigma, k}^L \in
   {}^*NDTM\right) \tag6.4 \\
   {}&\bigg[Mstate(x) = "yes" \wedge Mtime_y(x)
     \leq p\left(|x|\right)
   \bigg], \\
  {}&\left(\forall L\in {}^*NP\right)
  \left(\exists p\in {}^*\bar\Bbb NPOLY\right)
  \left(\exists M_{\delta , K, \Sigma, k}^L \in
   {}^*NDTM\right)
    \left(\forall y\in \left(\Sigma - \sqcup\right)^*\right)\\
   {}&\bigg[\left(Mstate(y) = "yes" \wedge Mtime_y(y)
     \leq p\left(|y|\right)\right) \to y\in L
   \bigg]. \\
\endalign
$$
\endproclaim

The $*$-transform property again shows that the internal Turing machines
has the same type of limitations as the classical Turing machines.
\proclaim{\bf Theorem 6.1} 
$P = NP$ if and only if
${}^*P = {}^*NP$ and 
$P \neq NP$ if and only if
${}^*P \neq {}^*NP$.
\endproclaim
\demo{Proof} This comes from the $*$-transform. \qed
\enddemo

\proclaim{Corollary 6.2}
Suppose $P \neq NP$, then there are internal languages
which can not be decided in $*$-polynomial time by internal
deterministic Turing machines.
\endproclaim
\demo{Proof} This follows from theorem 6.1. \qed
\enddemo

While the internal Turing machines has the same types of limitations
as the classical ones, the internal ones are much more powerful.
\proclaim{\bf Theorem 6.2} Let $L \in NP$ and 
$$
  {}^\sigma L = \left\{{}^*x| x \in L\right\},
  \tag6.5
$$
be the $*$-embedding of $L$. Then, 
there exists an internal
deterministic Turing machine ${}^*M$ such that
for all $x\in {}^\sigma L$,
on input $x$, 
${}^*M$ outputs $"yes"$ in $*$-polynomial time.
\endproclaim
\demo{Proof} Any language $L$ in $NP$ that is decided by
a nondeterministic Turing machine $N$ in polynomial time
$p(n)$ can be decided
by a deterministic Turing machine $M$ in time 
$\Cal O\left(c^{p(n)}\right)$, where $c > 1$ is
a constant depending on $N$ (see [4] and [8]).  
The theorem follows from Theorem 3.1 \qed
\enddemo

\comment
We will roughly sketch what the above means.
It is widely believed that $P \neq NP$, so we will
for the moment, assume that $P \neq NP$.
The set $3SAT$ consists of all boolean expressions
$\Phi$ in conjunctive normal form and all clauses
have 3 literals (see [4] and [8]).  
Thus, an element $\Phi \in 3SAT$ is given by
$$
  \Phi(u_1, \dots , u_m) = 
   \wedge_{i = 1}^n \Phi_i(u_1, \dots ,u_m) ,
$$
where $\Phi(u_1, \dots , u_m)$ and 
$\Phi_i(u_1, \dots , u_m)$ mean that
they are functions of $m$ boolean variables 
and the latter is a clause having 3 literals.
An internal element $\Phi$ of 
internal ${}^*3SAT$ is then given by
$$
  \Phi(u_1, \dots , u_m) = 
   \wedge_{i = 1}^n \Phi_i(u_1, \dots ,u_m) ,
$$
where $m, n\in {}^*\Bbb N$ and
again $\Phi_i(u_1, \dots ,u_m)$ is
a clause having 3 literals.  Thus,
internal ${}^*3SAT$ are allowed to have
infinite number of boolean variables 
and/or infinite number of 3 literal clauses.
The language $3SAT$ is $NP-complete$, meaning
that any language $L$ in $NP$ can be reduced
to the language $3SAT$ in polynomial time.
The $*$-transform of this gives any internal
language $L$ in ${}^*NP$ can be reduced to 
${}^*3SAT$ in $*$-polynomial time.  The
language ${}^*3SAT$ is the union of 
${}^\sigma 3SAT$ (the $*$-embedding of $3SAT$) and 
${}^*3SAT  - {}^\sigma 3SAT$.  Theorem 6.2
shows that 
${}^\sigma 3SAT$ can be decided by an internal
deterministic Turing machines in $*$-polynomial time.  Hence,
if $P \neq NP$, then 
${}^*3SAT  - {}^\sigma 3SAT$ contains
elements that can not be decided in $*$-polynomial time.
As far as the internal Turing machine goes, this offers
a separation between the easy and the hard problems
in ${}^*3SAT$.  This could be useful for obtaining
results in classical complexity theory but we must
be careful since the above two sets are external sets.
We will leave this for future research.
\endcomment


\Refs
\ref \no 1\by D. Aharonov
\paper Quantum Computation - A Review
\yr 1998
\jour Annual Review of Computational Physics, World Scientific, volume VI, ed. Dietrich Stauffer
\endref


\ref \no 2\by E. Bernstein and U. Vazirani
\paper Quantum Complexity Theory 
\yr 1997
\jour SIAM J. Comput. 26(5): 1411-1473
\endref

\ref \no 3\by N. Cutland
\book NonStandard Analysis and its Applications
\publ Cambridge University Press
\yr 1988
\endref

\ref \no 4\by D. Du and K. Ko
\book Theory of Computational Complexity
\publ John Wiley and Son
\yr 2000
\endref

\ref \no 5\by A. Hurd and P. Loeb
\book An Introduction to Nonstandard Real Analysis
\publ Academic Press
\yr 1985
\endref

\ref \no 6\by K. Ko
\book Complexity Theory of Real Functions
\publ Birkhauser
\yr 1991
\endref

\ref \no 7\by E. Nelson
\paper Internal Set Theory: A New Approach to Nonstandard Analysisj
\yr 1977
\jour Bulletin of American Mathematical Society, 83, 1165-1198
\endref

\ref \no 8\by C. Papadimitriou
\book Computational Complexity 
\publ Addison-Wesley
\yr 1995
\endref

\ref 
\no 9\by A. Robinson
\book Nonstandard Analysis
\publ Princeton Univ Pr; Revised edition
\yr 1996
\endref

\ref \no 10 
\by P. Shor
\paper Polynomial-Time Algorithms For Prime Factorization
and Discrete Logarithms on a Quantum Computer
\yr 1997
\jour SIAM J. Comput. 26(5): 1484-1509
\endref

\ref \no 11\by K. Stroyan and J. Luxemburg
\book Introduction to the Theory of Infinitesimals
\publ Academic Press
\yr 1976
\endref

\ref \no 12\by M. Nielsen and I. Chuang
\book Quantum Computation and Quantum Information
\publ Cambridge University Press
\yr 2000
\endref

\endRefs
\enddocument